\documentclass[twocolumn]{article}
\usepackage{cite}

\renewcommand\caption{}
\newcommand{\be}{\begin{equation}}
\newcommand{\ee}{\end{equation}}
\newcommand{\ba}{\begin{eqnarray}}
\newcommand{\ea}{\end{eqnarray}}

\def\beq{\begin{eqnarray}}
\def\eeq{\end{eqnarray}}
\def\ln{\,\mbox{ln}\,}

\def\ga{\gamma}

\def\la{\lambda}

\def\pa{\partial}

\begin{document}

\begin{center}
{\Large\sc on effective dimensional reduction in hyperbolic spaces} \vskip 6mm

{\small \bf E.V. Gorbar}
\footnote{E-mail address: gorbar@bitp.kiev.ua}

\vskip 6mm

{\small\sl \quad Bogolyubov Institute for Theoretical Physics, 03680 Kiev,
Ukraine}

\vskip 2mm
\end{center}

\vskip 6mm

\begin{abstract}

It is shown that the classical motion of massive particles in hyperbolic spaces
$H^D$ has a bounded character in $D-1$ coordinates. Studying the Dirac
equation, it is found that a bounded character of the classical motion
corresponds to the effective dimensional reduction $D+1 \to 1+1$ for fermions
in the infrared region in the quantum problem. This effective dimensional
reduction leads to the zero critical value of coupling constant for dynamical
symmetry breaking in hyperbolic spaces.

\end{abstract}

\vspace{5mm}

\section{Introduction}

\vspace{3mm}

It is well known that dynamical symmetry breaking (DSB) and a mass (gap)
generation for fermions usually require the presence of a strong attractive
interaction \cite{Fomin} in order to break symmetry that makes the quantative
study of DSB a difficult problem. Therefore, it is very interesting to consider
the cases where DSB takes place in the regime of weak coupling. Three such
examples are known.

The first is breaking of the $U(1)$ gauge symmetry in the presence of the Fermi
surface when some fermion states are filled. According to the
Bardeen--Cooper--Schrieffer theory of superconductivity \cite{BCS}, the Fermi
surface is crucial for the formation of a bound state and a symmetry breaking
condensate in the case of arbitrary small attraction between fermions. Indeed,
according to \cite{Shankar,Pol}, the renormalization group scaling in this case
is connected only with the direction perpendicular to the Fermi surface.
Therefore, the effective dimension of spacetime is $1 + 1$ from the viewpoint
of renormalization group scaling. Since a bound state can be generated for any
attraction in $1 + 1$ dimension, this implies that the critical coupling
constant is zero in this case.

The other example of DSB in the regime of weak coupling is DSB in a constant
magnetic field \cite{GMSh1,GMSh2}, where chiral symmetry is also dynamically
broken for an arbitrary weak attraction. The physical reason for this is the
effective dimensional reduction of spacetime for fermions in the infrared
region by 2 units in a constant magnetic field (DSB in a magnetic field in
spacetimes of dimension higher than four was considered in \cite{G1}). Indeed,
electrons being charged particles cannot propagate in directions perpendicular
to the magnetic field when their energy is much less than the Landau gap
$\sqrt{|eB|}$. This leads to the effective dimensional reduction in the
infrared region by 2 units.

The dynamics of fermions in hyperbolic spaces $H^D$ gives the third known
example of DSB with zero critical coupling constant \cite{C1,C2,C3,C4,Ina}.
Analysing the heat kernel in spacetimes $R \times H^D$, it was shown in
\cite{G2} that the zero value of critical coupling constant for DSB in this
case is connected also with the effective dimensional reduction $D+1 \to 1+1$
for fermions in the infrared region. In a recent paper \cite{Ebert}, chiral and
diquark condensates were studied in the extended NJL model in the hyperbolic
space $H^3$ and it was shown that negative curvature increases the values of
both condensates.

The dynamics of quantum fields in hyperbolic spaces has received a great deal
of attention in the literature (see, e.g., \cite{BCVZ, MS} and the references
therein). Callan and Wilczek advocated hyperbolic spaces as a geometric means
to regularize the infrared divergences of quantum field theories \cite{CW}.
Further, spatial sections of open Friedmann-Robertson-Walker models are
hyperbolic spaces. The near-horizon optical metric of non-extreme black holes
is asymptotically hyperbolic also \cite{N1,N2,N3}.

Although the analysis of the heat kernel in \cite{G2} clearly showed the
presence of the effective dimensional reduction in spacetimes $R \times H^D$,
physical reasons for such a reduction remained unknown because the heat kernel
is only an integral characteristics of the dynamics of the system. Therefore,
it would be certainly desirable to clarify physical reasons of such a
reduction. Fortunately, there exists a physically transparent way to
demonstrate the occurence of the effective dimensional reduction in $H^D$.
According to \cite{APNY}, the effective dimensional reduction in the infrared
region takes place in a quantum problem only when classical motion in the
corresponding problem has a bounded character with respect to the coordinates
over which the dimensional reduction occurs. Let us consider, for example, the
effective dimensional reduction in a constant magnetic field. In this case, a
charged classical particle moves on circular orbits in the plane perpendicular
to the constant magnetic field. Since the radius of orbit is proportional to
energy, the particle can go to infinity only if it has infinite energy.
Therefore, a charged particle of finite energy always moves only in a finite
region of the plane perpendicular to the constant magnetic field. This bounded
character of motion means that the system is effectively of a finite size and
leads to the effective dimensional reduction by 2 units in the infrared region
in the quantum problem \cite{GMSh1}. Therefore, it is interesting to consider
the classical motion in hyperbolic spaces and see whether this motion has
indeed a bounded character with respect to the coordinates over which the
effective dimensional reduction takes place. Briefly this motion and an
analysis of the Dirac equation in hyperbolic spaces were considered in a recent
paper \cite{Lobachevsky}. In the present paper, we consider these problems in
more detail.

\section{Classical motion in hyperbolic spaces}

In this section, we study the classical motion in hyperbolic spaces.  Let us
first recall what hyperbolic spaces are and introduce the necessary notation.
Hyperbolic spaces $H^D$ are symmetric Riemannian spaces of constant negative
curvature whose interval in the Poincar{\'e} coordinates is given by \beq dl^2
= \frac{a^2}{x_1^2}(dx_1^2 + dx_2^2 + ... + dx_D^2), \label{HDmetric} \eeq
where $x_1 > 0$ and $a$ is the curvature radius. Further, the spacetime that we
consider is $R \times H^D$ and the corresponding spacetime interval is
\begin{equation}
ds^2=c^2dt^2-dl^2.
\label{spacetime-interval}
\end{equation}
Relativistic motion of free particles proceeds along geodesics (see, e.g.,
\cite{course-2}) which are extrema of the action
\begin{equation}
S=-mc\int ds = -mc^2\int
\sqrt{1-\sum_{k=1}^{D}\frac{a^2\dot{x}_k^2}{c^2x_1^2}}\,\,dt.
\label{action}
\end{equation}
Since the Lagrangian in (\ref{action}) does not depend expli\-citly on $t$,
energy
\begin{equation}
E=\sum_k p_k\dot{x}_k -
L=\frac{mc^2}{\sqrt{1-\sum_k\frac{a^2\dot{x}_k^2}{c^2x_1^2}}}
\label{energy}
\end{equation}
is an integral of motion. Using this fact, it is not difficult to find the
equations of motion. They are
\begin{equation}
\frac{d^2 x_1}{dt^2} =
\frac{\dot{x_1}^2-\dot{x}_2^2-\dot{x}_3^2-...-\dot{x}_D^2}{x_1}\,,
\label{eq-motion}
\end{equation}
\begin{equation}
\frac{d}{dt}(\frac{\dot{x}_2}{x_1^2})=...=\frac{d}{dt}(\frac{\dot{x}_D}{x_1^2})=0\,.
\label{new-form}
\end{equation}
Eqs.(\ref{new-form}) are easily integrated
\begin{equation}
\dot{x}_2=C_2x_1^2\,,\,\,\,.\,.\,.\,\,,\,\,\,\dot{x}_D=C_Dx_1^2,
\label{space-solutions}
\end{equation}
where $C_2\,,\,.\,.\,.\,,\,C_D$ are integration constants. Substituting
(\ref{space-solutions}) in (\ref{eq-motion}), the equation for $x_1$ takes the
form
\begin{equation}
\frac{d^2 x_1}{dt^2} - \frac{\dot{x}_1^2}{x_1} = - C^2x_1^3,
\label{x-1-equation}
\end{equation}
where $C^2=C_2^2+...+C_D^2$. Making the change of variable $z=\ln x_1$ (note
that this change of variable is unambiguous because $x_1 > 0$), we find the
following equation for $z$:
\begin{equation}
\frac{d^2 z}{dt^2} = - C^2 e^{2z}.
\label{nonlinear-equation}
\end{equation}

Since this equation does not depend explicitly on $t$, we introduce
$y(z)=\dot{z}$ and obtain the equation
\begin{equation}
\frac{d\,y^2}{dz}=-2C^2e^{2z}, \label{u-equation}
\end{equation}
which is easily integrated
\begin{equation}
y^2=-C^2e^{2z}+A, \label{u-solution}
\end{equation}
where $A$ is an integration constant. Since $y^2$ is positive, the right-hand
side of Eq.(\ref{u-solution}) should be positive also, therefore, $A>0$. Since
$y=\dot{z}$, for $A=u^2$, it follows from Eq.(\ref{u-solution}) that
\begin{equation}
\dot{z}=\pm\sqrt{u^2-C^2e^{2z}}.
\label{first-order}
\end{equation}
Without the loss of generality, we assume the sign minus in
Eq.(\ref{first-order}) because in view of the change $t \to -t$ both sings are
equivalent. Then integrating Eq.(\ref{first-order}), we obtain the sought
solution
\begin{equation}
z(t)=\ln \frac{u}{C\cosh (ut+b)},
\label{z-solution}
\end{equation}
where $b$ is an integration constant. Using (\ref{z-solution}), it is not
difficult to integrate Eqs.(\ref{space-solutions}) and get the law of motion in
other $D-1$ space coordinates $x_2,...,x_D$. Thus, we find the following
classical trajectories of motion in $H^D$ (geodesics):
\begin{eqnarray}
z(t) = \ln(x_1(t)) = \ln\frac{u}{C\cosh(ut + b)}, \nonumber \\
x_2(t) = \frac{uC_2}{C^2} \tanh(ut + b) + \tilde{C}_2, \nonumber \\
. \quad\quad\quad\quad\quad\quad\quad\quad\quad \nonumber \\
. \quad\quad\quad\quad\quad\quad\quad\quad\quad \nonumber \\
. \quad\quad\quad\quad\quad\quad\quad\quad\quad \nonumber \\
x_D(t)  = \frac{uC_D}{C^2} \tanh(ut + b) + \tilde{C}_D,
\label{trajectories}
\end{eqnarray}
where $C=\sqrt{C_2^2+C_3^2+...+C_D^2}$ and $u, b, C_2, \tilde{C}_2,...,$
$C_D,\tilde{C}_D$ are arbitrary constants (there are 2D of them and, for a
particular trajectory, they are fixed by initial conditions). Finally, it is
not difficult to check that energy (\ref{energy}) that corresponds to this
motion equals $E=mc^2/(1-v^2/c^2)^{1/2}$, where $v=au$. This means that $u$
cannot exceed $c/a$.

At this point, we would like to comment why we prefer to work with the $z=\ln
x_1$ coordinate rather than $x_1$. In a certain sense, the $z$ coordinate is
more natural from the viewpoint of metric (\ref{HDmetric}) because then the
corresponding contribution to the interval has the flat space form $dz^2$ and
$z$ takes values on the whole real axis unlike the $x_1$ coordinate which takes
values only on the positive semiaxis. It is easy to see from Eqs.
(\ref{trajectories}) that the motion in the coordinate $z$ has the same
character as the usual flat space motion except a time interval of order $1/u$.
Indeed, according to Eqs.(\ref{trajectories}), the classical particle moves
like $z(t)=ut + z_1$ for $t \ll -\frac{1}{u}$. For $|t| \le \frac{1}{u}$, its
motion differs from the familiar inertial motion in flat space. For $t \gg
\frac{1}{u}$, the particle moves like $z(t)=-ut+z_2$, i.e., it changes the
direction of motion and goes back to $-\infty$, where it started its motion.
Thus, the motion in the coordinate $z$ has a familiar inertial character except
a time interval of order $1/u$ with the resulting change of the direction of
motion.

On the other hand, according to Eqs.(\ref{trajectories}), motion in
$x_2,...,x_D$ coordinates has a completely different character. The particle is
practically motionless for almost all period of time except the time interval
of order $1/u$ when it moves some finite distance. Therefore, the classical
motion in these coordinates has a bounded character. On the other hand,
coordinates can be arbitrarily chosen in curved spacetime, therefore, we should
search for an invariant way to demonstrate a bounded character of the classical
motion in hyperbolic spaces. For this, we consider Killing vector fields
connected with translations in $x_2,\,x_3,...,\,x_D$ coordinates.

The hyperbolic space has $D-1$ ``translational'' Killing vector fields $\xi_i$
($i = \overline{2, D}$), which, in the coordinates employed in this paper, are
given by
\begin{equation}
\xi_i = {\partial \over
\partial x_i} \, , \quad i = 2, \ldots , D \, .
\label{vector-fields}
\end{equation}
Note that these Killing vector fields realize translations in the corresponding
coordinates and form a commuting subalgebra of Killing vector fields in the
hyperbolic space. Using the normalized vector fields
\begin{equation}
n_i = { \xi_i \over \left( - \xi_i \cdot \xi_i \right)^{1/2}} \, ,
\label{normalized}
\end{equation}
where the central dot denotes the scalar pro\-duct in metric (\ref{HDmetric}),
we calculate the follo\-wing reparametrization-invariant integral along a
geodesic:
\begin{equation}
l^{(geo)}_{i}  = \int |\left( u_t \cdot n_i \right)| dt =
 a \int \sqrt{\frac{\dot{x_i}^2}{x_1^2}}
dt = \frac{\pi a|C_i|}{C}\,,
\label{finite}
\end{equation}
where $u_t$ is the tangent vector to the geodesic and integration in $t$ goes
from $-\infty$ to $+\infty$. Clearly, Eq.(\ref{finite}) describes the geodesic
motion in $x_i$ ($i=\overline{2,D}$) coordinates and, what is important, it
does not depend on the choice of coordinates because $n_i$ are expressed
through Killing vector fields. Since $|C_i|<C$, the quantities $l^{(geo)}_{i}$
are always less than $L=\pi a$. This is an important result. First of all, note
that if we fix by hand $x_1$ and integrate in Eq.(\ref{finite}) over $d x_i$
(of course, the corresponding curve is not a geodesics), then the integral in
(\ref{finite}) will be infinite. Therefore, a priori, one may expect the space
$l^{(geo)}_{2}\times ... \times l^{(geo)}_{D}$ to be a subspace of
$(D-1)$-dimensional Euclidean space $R^{D-1}$ unbounded in all $D-1$
directions. However, according to Eq.(\ref{finite}), this is not correct and
the space $l_{2}^{(geo)}\times ... \times l_{D}^{(geo)}$ connected with
geodesic motions is a $(D-1)$-dimensional cube $L^{D-1}$. Since $l^{(geo)}_{i}$
are invariant with respect to the change of coordinates, we conclude that
geodesic motion in hyperbolic spaces has a bounded character in $D-1$
directions. Finally, let us emphasize that this result does not mean that
geodesic motion takes place in a certain finite region in $x_2,...,x_D$
coordinates. Eqs.(\ref{trajectories}) imply, for example, that one can connect
any two points with diffe\-rent $x_i$ by a geodesics because $|C_i|/C^2$ can be
arbit\-rary large unlike $|C_i|/C$ that enters Eq.(\ref{finite}), which is
always bounded by unity.

At this point, it is instructive to calculate the spatial interval of a
geodesics. Obviously, it is invariant with respect to coordinate
transformations that do not involve the time coordinate. Using
Eqs.(\ref{HDmetric}) and (\ref{trajectories}), we find
\begin{equation}
l^{(geo)} = \int_{t_a}^{t_b}
\sqrt{\sum_{k=1}^{D}\frac{a^2\dot{x_k}^2}{x_1^2}}\,\,dt = ua \int_{t_a}^{t_b}
dt.
\label{full-interval}
\end{equation}
Obviously, it diverges as $t_a$ or $t_b$ tends to infinity. Finally, note that
we can consider $l_{1}^{(geo)}=a \int \sqrt{\frac{\dot{x_1}^2}{x_1^2}} dt$,
which is an analog of (\ref{finite}) for motion in the $x_1$ coordinate and
which, like $l^{(geo)}$, diverges. Using $\zeta_1=\partial_{x_1}$, this
integral can be represented also as
\begin{equation}
l^{(geo)}_{1}  = \int |\left( u_t \cdot n_1 \right)| dt\,,
\label{partial-interval}
\end{equation}
where $n_1=\zeta_1/(-\zeta_1 \cdot \zeta_1)^{1/2}$ is the normalized vector
field which is orthogonal to the vector fields $n_i$ ($i = \overline{2, D}$).
Although $\zeta_1$ is similar to $\xi_i$ given by Eq.(\ref{vector-fields}), it
is not a Killing vector field in hyperbolic spaces. Nevertheless, being
orthogonal to all $\xi_i$ vector fields, it is, in fact, uniquely defined.
Therefore, $l^{(geo)}_1$ has a clear geometric meaning. Note that the vector
fields $\zeta_1$ and $\xi_i$ form a commuting algebra of vector fields
equivalent to the algebra of ``translational'' Killing vector fields of
$D$-dimensional Euclidean space. Therefore, perhaps, the most clear signature
of a bounded character of motion in hyperbolic spaces is provided by the ranges
of values of $l^{(geo)}_1$ and $l^{(geo)}_i$.

\section{Solutions of the Dirac equation and the effective dimensional
reduction}

In the previous section, we showed that the classical motion in hyperbolic
spaces has a bounded character in $D-1$ coordinates. According to \cite{APNY},
this should lead to the effective dimensional reduction $D + 1 \to 1 + 1$ in
the quantum problem. In order to show this, we consider in this section
solutions of the Dirac equation in hyperbolic spaces (see also \cite{BCVZ,N2}
where different approaches to the derivation of solutions are used).

The Dirac equation in hyperbolic spaces $H^D$ has the form
$$
\left(i\ga^0\pa_t + \frac{ix_1}{a}\ga^1\pa_1 + ... + \frac{ix_1}{a}\ga^D\pa_D
\right.
$$
\begin{equation}
\left. - \frac{i(D-1)}{2a}\ga^1 - m\right)\psi = 0.
\label{HDDequation}
\end{equation}
In order to solve this equation, we multiply it by $i\hat{D} + m$ and obtain
the following second order differential equation:
$$
(-\pa_t^2 + \frac{(D-1)^2}{4a^2} - \frac{D-2}{a^2} x_1\pa_1
$$
$$
+ \frac{x_1^2}{a^2}(\pa_1^2 + ... + \pa_D^2) - \frac{x_1}{a^2}\ga^1\ga^2\pa_2 -
$$
\begin{equation}
\frac{x_1}{a^2}\ga^1\ga^3\pa_3 - ... - \frac{x_1}{a^2}\ga^1\ga^D\pa_D -
m^2)\psi = 0.
\label{DEsquared}
\end{equation}
Obviously, we can seek solution in the form $\psi=e^{-iEt + ip_2x_2 + ... +
ip_Dx_D}f(x_1)$. Then we have
$$
(E^2 + \frac{(D-1)^2}{4a^2} - \frac{D-2}{a^2} x_1\pa_1 +
\frac{x_1^2\pa_1^2}{a^2}
$$
$$
- \frac{x_1^2}{a^2}(p_2^2 + ... + p_D^2) -i\frac{x_1}{a^2}\ga^1\ga^2p_2
$$
\begin{equation}
- i\frac{x_1}{a^2}\ga^1\ga^3p_3 - ... - i\frac{x_1}{a^2}\ga^1\ga^Dp_D - m^2
)f(x) = 0.
\end{equation}
Further, the Dirac $\gamma$ matrices are present in this equation only in the
operator
\begin{eqnarray*}
A=-i\frac{x_1}{a^2}\ga^1\ga^2p_2 - i\frac{x_1}{a^2}\ga^1\ga^3p_3 - ... -
i\frac{x_1}{a^2}\ga^1\ga^Dp_D.
\end{eqnarray*}
Since this is a Hermitian operator and its square
\begin{eqnarray*}
A^2=\frac{x_1^2}{a^4}(p_2^2 + ... + p_D^2)
\end{eqnarray*}
is a unit matrix, it can be diagonalized and obviously its eigenvalues are
equal to
\begin{equation}
\sigma\frac{x_1}{a^2}\sqrt{p_2^2 + ... + p_D^2},
\label{eigenvalues}
\end{equation}
where $\sigma=\pm$. Making the change of variable $x=x_1\sqrt{p_2^2+ ... +
p_D^2}$, we obtain
$$
\left(E^2 + \frac{x^2}{a^2}(-1+\pa_x^2) + \frac{(D-1)^2}{4a^2} \right.
$$
\begin{equation}
\left. - \frac{(D-2)x}{a^2}\pa_x + \frac{\sigma x}{a^2} - m^2 \right)f(x) = 0.
\label{effectiveDE}
\end{equation}
The absence of any dependence on $p_2,...,p_D$ in this equation is remarkable
because it means that energy does not depend on them, i.e., energy is the same
for any $p_2,..., p_D$. Eq.(\ref{effectiveDE}) has the form of equation of an
$(1 + 1)$-dimensional problem and it is not difficult to find its spectrum $E =
\pm \sqrt{\nu^2/a^2 + m^2}$, where $\nu$ takes values in $(0, +\infty)$. The
effective (1 + 1)-dimensional form (\ref{effectiveDE}) of the Dirac equation
means the effective dimensional reduction of the dynamics of fermions in
hyperbolic spaces. From the mathematical viewpoint, this reduction is related
to the spherical and scale symmetries of the $H^D$ metric written in the
Poincar{\'e} coordinates. Indeed, the spherical symmetry of the $x_2,...,x_D$
part of metric (\ref{HDmetric}) reduces the dependence of energy on
$p_2,...,p_D$ to the dependence on only one invariant $p^2=p_2^2+...+p_D^2$ and
then the symmetry of metric (\ref{HDmetric}) with respect to scale
transformations $x_k \to \la x_k\, (k=\overline{1,D})$ (recall the change of
variable $x=x_1\sqrt{p_2^2+ ... + p_D^2}$ that we performed after
Eq.(\ref{eigenvalues})) eliminates any dependence on $p_2,...,p_D$ in Eq.
(\ref{effectiveDE}) for eigenfunctions.

\section{Conclusion}

Studying the classical motion of massive particles and solving the Dirac
equation in hyperbolic spaces $H^D$, we clarified the physical reasons for the
effective dimensional reduction $D+1 \to 1+1$ for fermions in the infrared
region in hyperbolic spaces.

We showed that the classical motion has a bounded character in $D-1$
coordinates and, according to Eq.(\ref{finite}), the physical system is
effectively of a finite size with respect to these coordinates. On the other
hand, the classical motion in the $z=\ln x_1$ coordinate has a familiar
inertial character except a finite interval of time when the particle changes
its direction of motion and goes back to $-\infty$, where it started its
motion. Therefore, for any classical trajectory with nonzero velocity, motion
in the $x_1$ coordinate ensures that $l^{(geo)}_1$ is infinite unlike
$l_i^{(geo)}$ which describe geodesic motions in other $D-1$ coordinates.

Solving the Dirac equation in spacetimes $R \times H^D$, we showed that due to
the spherical and scale symmetries the initial Dirac problem is reduced to an
effective $(1 + 1)$-dimensional problem. This result agrees with the
conclusions of \cite{APNY} that the effective dimensional reduction in a
quantum problem is connected with a bounded character of the classical motion
with respect to the coordinates over which the reduction takes place.

\vspace{4mm}

\centerline{\bf Acknowledgements}

\vspace{2mm}

The author is grateful to V.P. Gusynin, V.A. Miransky, and Yu.V. Shtanov for
useful remarks and suggestions. The author especially thanks Yu.V. Shtanov for
the suggestion to use Killing vector fields in order to demonstrate in an
invariant way a bounded character of motion in hyperbolic spaces. This work was
supported by the "Cosmomicrophysics" program, by the State Foundation for
Fundamental Research under the grant F/16-457-2007, and by the Program of
Fundamental Research of the Physics and Astronomy Division of the National
Academy of Sciences of Ukraine.

\end{document}